\documentclass[manuscript]{acmart}

\usepackage{graphicx}
\usepackage{caption}
\usepackage{booktabs}
\usepackage{subcaption}
\usepackage{array}
\usepackage{float}
\usepackage{makecell}
\usepackage{mathtools, amsmath}
\usepackage{relsize}
\usepackage{lscape}
\usepackage{multicol, multirow}
\usepackage{enumitem}
\usepackage{tablefootnote}
\usepackage{xcolor}
\usepackage{array}
\usepackage{amsmath,amsfonts}
\usepackage{algorithmic}
\usepackage{algorithm}
\AtBeginDocument{%
  }

\copyrightyear{2025}
\acmYear{2025}
\setcopyright{rightsretained}
\acmConference[CHI EA '25]{Extended Abstracts of the CHI Conference on Human Factors in Computing Systems}{April 26-May 1, 2025}{Yokohama, Japan}
\acmBooktitle{Extended Abstracts of the CHI Conference on Human Factors in Computing Systems (CHI EA '25), April 26-May 1, 2025, Yokohama, Japan}\acmDOI{10.1145/3706599.3720197}
\acmISBN{979-8-4007-1395-8/2025/04}

\begin{document}


\title[Wearable meets LLMs for Stress Management: A Duoethnographic Study]{Wearable Meets LLM for Stress Management: A Duoethnographic Study Integrating Wearable-Triggered Stressors and LLM Chatbots for Personalized Interventions}
\titlenote{A version of this work appears in ACM CHI 2025.}




\author{Sameer Neupane}
\authornote{Corresponding Author}

\authornote{Both authors contributed equally to this research.}

\email{sameer.neupane@memphis.edu}
\affiliation{%
  \institution{University of Memphis}
  \streetaddress{}
  \city{Memphis}
  \state{Tennessee}
  \country{USA}
  \postcode{}
}

\author{Poorvesh Dongre}
\authornotemark[3]
\email{poorvesh@vt.edu}
\affiliation{%
  \institution{Virginia Tech}
  \streetaddress{}
  \city{Blacksburg}
  \state{Virginia}
  \country{USA}
  \postcode{}
}

\author{Denis Gracanin}

\email{gracanin@vt.edu}
\affiliation{%
  \institution{Virginia Tech}
  \streetaddress{}
  \city{Blacksburg}
  \state{Virginia}
  \country{USA}
  \postcode{}
}

\author{Santosh Kumar}

\email{santosh.kumar@memphis.edu}
\affiliation{%
  \institution{University of Memphis}
  \streetaddress{}
  \city{Memphis}
  \state{Tennessee}
  \country{USA}
  \postcode{}
}

\renewcommand{\shortauthors}{Sameer, et al.}

\begin{abstract}
We use a duoethnographic approach to study how wearable-integrated LLM chatbots can assist with personalized stress management, addressing the growing need for immediacy and tailored interventions. 
Two researchers interacted with custom chatbots over 22 days, responding to wearable-detected physiological prompts, recording stressor phrases, and using them to seek tailored interventions from their LLM-powered chatbots.
They recorded their experiences in autoethnographic diaries and analyzed them during weekly discussions, focusing on the relevance, clarity, and impact of chatbot-generated interventions. Results showed that even though most events triggered by the wearable were meaningful, only one in five warranted an intervention. It also showed that interventions tailored with brief event descriptions were more effective than generic ones. 
By examining the intersection of wearables and LLM, this research contributes to developing more effective, user-centric mental health tools for real-time stress relief and behavior change.

\end{abstract}




\maketitle

\section{Introduction}

Stress is defined as a state of mental or emotional strain resulting from adverse or demanding circumstances. 
Stress has become a pervasive challenge in modern life, with chronic stress posing significant risks to mental and physical health \cite{sapolsky2004zebras,chrousos2009stress,dai2020chronic, guilliams2010chronic, leonard2015multi}. 
This issue holds considerable significance in Human-Computer Interaction (HCI), as technology plays an increasingly central role in daily life and, consequently, in mental health interventions.

Traditional approaches to stress management, such as therapy and structured programs, are valuable but often constrained by accessibility, cost, and the lack of immediacy. 
Wearable devices have emerged as promising tools for stress tracking, offering real-time monitoring of physiological stress indicators, such as heart rate variability, to provide timely insights into stress levels \cite{Staff_2022,garmin-watch,Kuzmowycz_2023}. 
Complementing this, generative AI has been used to develop on-demand mental health support chatbots \cite{liu2023chatcounselor,lai2023psy}. 
While LLM-driven AI solutions have shown promise in supporting mental health diagnosis~\cite{nie2024llm} and providing mental health first aid~\cite{ji2024mindguard}, they often exhibit behaviors typical of low-quality therapy~\cite{chiu2024computational}. 
Recent LLM-powered journaling approaches integrating passively collected behavioral patterns have enhanced self-awareness and well-being~\cite{nepal2024mindscape}. 
However, passive smartphone data may miss momentary events (e.g., anxiety, frustration) that wearables may catch from physiological arousal. 
In both cases, not all detected events may need an intervention. 
We envision a future where wearables, smartphones, and generative AI work together to identify stressors automatically, determine which stressors require intervention, and provide tailored interventions.

In this paper, we take the first steps by investigating the opportunities and challenges in integrating wearable-triggered stressor journaling with LLM chatbots to provide personalized interventions. 
We employed a duoethnographic methodology, a collaborative autoethnography involving two researchers, to capture detailed, reflexive accounts of integrating these technologies. 
In a 22-day study, two PhD students (who also experience high self-reported stress) used CuesHub \cite{CuesHub}, a smartwatch-based system offering real-time wearable triggered stressor journaling, while simultaneously using a personalized LLM-based stress intervention chatbot built with OpenAI’s custom GPT. 
Each researcher engineered prompt templates tailored to their individual stress triggers, coping styles, and daily routines, following a lightweight framework informed by established stress management strategies. 
This personalized approach allowed a more nuanced exploration of how wearables and LLM chatbot interventions can be adapted to user needs and preferences.

Our motivation lies in harnessing the complementary strengths of wearables and LLM chatbots for an integrated stress management solution. 
Our research focuses on two questions: (1) How often do users seek interventions for momentary events triggered by wearables? and (2) How does integrating very brief descriptions of momentary events into LLMs enhance stress management interventions?
By documenting our lived experiences, we aim to inform the design of multi-modal AI solutions that can track stressors in real time, offer interventions, and adapt to the complexities of individual user needs. 
Our final goal is to articulate design considerations and insights that can guide broader HCI efforts to develop integrated AI systems for stress management.

\section{Background and Related Works}
Technology-powered interventions have become increasingly popular for supporting mental health and well-being. 
They provide real-time data, adapt to personal needs, and reduce barriers such as cost and accessibility.

\subsection{Wearables for Stress Tracking, Stressor Journaling, and Interventions}
Wearable devices have emerged as one of the most widely used AI-driven tools for recognizing affect states, including stress from physiological signals. 
During the past two decades, the scientific community has made significant advances in stress detection, using wearable technology to monitor physiological responses in real-time indicative of stress~\cite{wong-2019-stress-empatica, ollander-2016-stress-empatica, gjoreski2016continuous}. 
Building on this foundation, commercial devices such as Fitbit~\cite{Staff_2022}, Garmin~\cite{garmin-watch}, Whoop~\cite{Kuzmowycz_2023}, and Empatica Embrace \cite{CiteDrive2022_1} have recently integrated stress tracking capabilities. 
Passively collected data has been leveraged to create innovative visualizations for self-reflection~\cite{sanches2010mind,kocielnik2015personalized,kocielnik2013smart,stepanovic2019designing,xue2019affectivewall}. 
Furthermore, passive stress detection has effectively delivered real-time interventions, addressing stress precisely when it occurs~\cite{howe2022design,battalio2021sense2stop}. 
However, participants in these studies have expressed the desire for interventions to match the source of stress. 
Recent work has shown how stress detection can be used to collect stressors and how incorporating stressors into self-reflective visualizations can lead to behavioral changes~\cite{neupane2024momentary}. 
This study takes the next step of integrating stressors with chatbot interactions for personalized stress interventions.

\subsection{LLM for Mental Health Support}
Mental health support chatbots trace their origins back to ELIZA, one of the first computer programs to simulate human conversation \cite{weizenbaum1966eliza}. 
Since then, several rule-based and semi-automated chatbots have emerged that offer psycho-education, basic coping strategies, and limited therapeutic interactions. 
However, the emergence of LLMs has transformed the development of conversational agents and created novel opportunities to provide users with advanced mental health chatbots. 
Liu et al. \cite{liu2023chatcounselor} developed an LLM chatbot, ChatCounselor, by fine-tuning an open-source pre-trained LLM on a dataset prepared from 260 in-depth interviews between patients and psychologists. 
The LLM chatbot by Lai et al. \cite{lai2023psy} was developed by fine-tuning two Chinese pre-trained LLMs with real-world professional Q\&A datasets from psychologists and psychological articles to develop their mental health chatbot. 
Dongre et al. \cite{dongre2024integrating, dongre2024physiology} also fine-tuned an open-source LLM on a Q\&A dataset scraped from a mental health website to develop their LLM chatbot for stress management.   
CaiTI, a Conversational AI Therapist, uses LLMs, smart devices, and reinforcement learning to provide personalized psychotherapeutic interventions, enhancing mental health self-care~\cite{nie2024llm}. 
BOLT, a framework for evaluating LLM therapists’ conversational behavior, highlights the need for improvements to achieve high-quality care~\cite{chiu2024computational}. 
MindScape introduces an AI-powered journaling approach that integrates behavioral patterns like sleep, location, and engagement to deliver personalized, context-aware prompts~\cite{nepal2024mindscape}.
These works show the vast potential of LLM to power tailored interventions on demand. 
The novelty of this study lies in its integration with wearable AI to deliver automated, personalized, and context-aware stress interventions.

\begin{figure*}[t]
     \centering
     \begin{subfigure}[b]{0.41\textwidth}
         \centering
         \includegraphics[width=\textwidth]{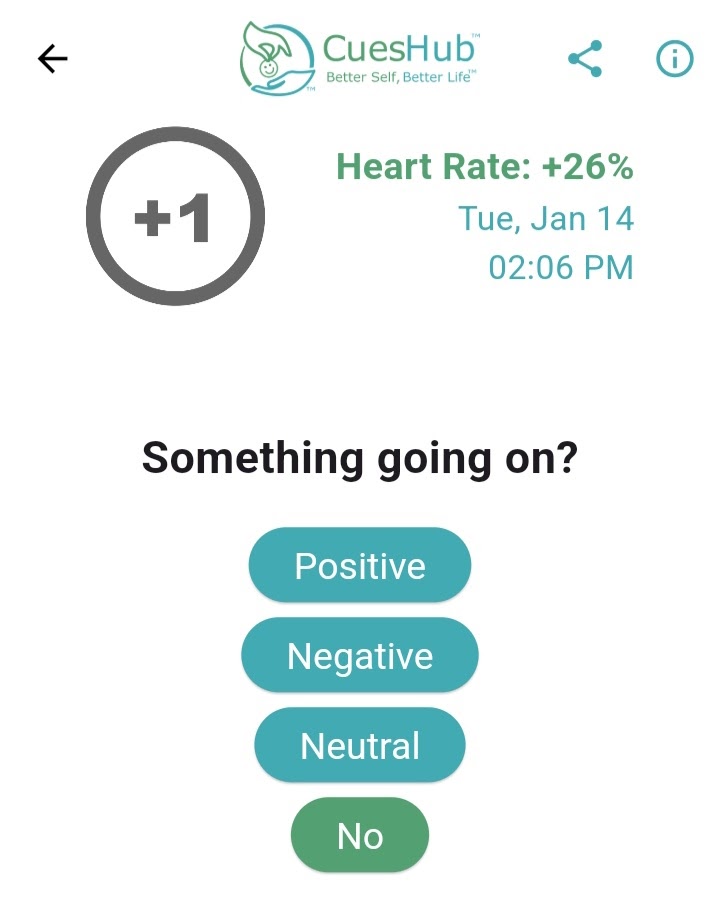}
\caption{Valence rating screen of the prompted event}
     \label{fig:Cueshub_1}
     \end{subfigure}
     \hfill
     \begin{subfigure}[b]{0.4\textwidth}
         \centering
         \includegraphics[width=\textwidth]{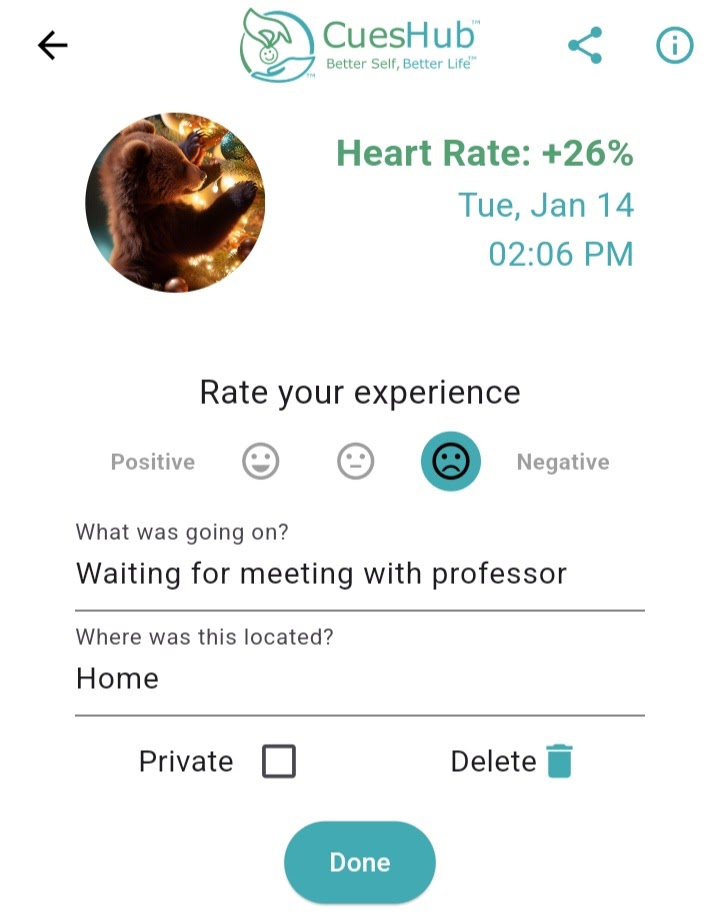}
\caption{Event description screen of the prompted event}

         \label{fig:Cueshub_2}
         \captionsetup{justification=centering}
         
     \end{subfigure}
      \caption{CuesHub app screenshots for valence and event descriptions}
        \label{fig:cueshub}
        \Description{Figure shows the screenshots of the CuesHub app to rating of the prompted events}
\end{figure*}

\section{Methods}
\subsection{Duoethnography}
This study employs a \textit{duoethnographic approach}, a qualitative method exploring how two or more researchers interpret and give meaning to shared experiences~\cite{cifor2020gendered,sawyer2012duoethnography}. 
It builds on autoethnographic methods~\cite{ellis2011autoethnography,creswell2016qualitative,vakeva2024disorientation,o2014gaining}, which have been applied in diverse contexts, including the use of LLM chatbots for thesis writing, creative expression, and self-identity~\cite{schwenke2023potentials, ghajargar2022redhead, olasik2023good}.
For this study, two researchers independently interacted with the wearables and personalized LLM chatbots for stress management, documenting their experiences and reflections in autoethnographic diaries. 
This dual autoethnographic approach provided introspective data on how integrating the two AI systems influenced perceived effectiveness in managing everyday stress.
Weekly meetings between the researchers enabled shared reflection, discussion, and comparative analysis of individual perspectives. 
The study’s insights into integrating wearables and LLM chatbots for stress management were enriched by combining individual depth with collaborative breadth.

\subsection{Wearable for Stressor Journaling}
Both researchers wore Samsung Galaxy Watch 6 devices equipped with the CuesHub smartwatch app~\cite{CuesHub}, serving as the wearable AI platform for detecting physiological events associated with stress. Upon detecting a physiological event, the smartwatch app prompted researchers via their smartphones, requesting them to provide brief descriptions of the event. 
Researchers rated the events using one of four options: \emph{Positive}, \emph{Negative}, \emph{Neutral}, or \emph{No} (see Figure~\ref{fig:Cueshub_1}). For responses indicating \emph{Positive}, \emph{Negative}, or \emph{Neutral}, researchers were further prompted to answer two questions: \emph{What was going on?} and \emph{Where was this located?} (see Figure~\ref{fig:Cueshub_2}).

\subsection{Custom GPT for Stress Intervention}
The chatbots employed in this study used OpenAI’s GPT-4o~\footnote{\url{https://openai.com/index/gpt-4/}} as their underlying language model. 
To investigate the effectiveness of different interaction paradigms, we explored two contrasting approaches to chatbot design, emphasizing distinct user experiences and intervention delivery styles. 
\textit{Researcher A ($R_A$)} designed their chatbot, named \emph{DeStressify}, to emulate a zero-shot prompting approach akin to intervention systems that automatically deliver stress interventions upon detecting a user’s stress\cite{howe2022design,battalio2021sense2stop}. 
When experiencing the need for intervention, \emph{Researcher A} prompted the chatbot with their stressor and location. 
Most interactions were transactional, consisting of a single prompt-response exchange where the chatbot provided an intervention in the traditional instructive manner often used in text messaging.

\emph{Researcher B ($R_B$)}, on the other hand, adopted the modern approach for their chatbot, named \emph{StressGPT}. 
This chatbot was structured to facilitate a more therapy-like interaction, allowing them to engage in dynamic, multi-turn dialogues. 
This conversational style aimed to simulate collaborative interactions, with \emph{StressGPT} mimicking personalized coaching sessions by refining suggestions or providing layered support through iterative exchanges, enabling deeper reflection and tailored advice.
Details about the chatbot designs for both researchers are provided in Appendix~\ref{sec:design_templates}. 

\begin{figure*}[t]
     \centering
     \begin{subfigure}[b]{0.47\textwidth}
         \centering
         \includegraphics[width=\textwidth]{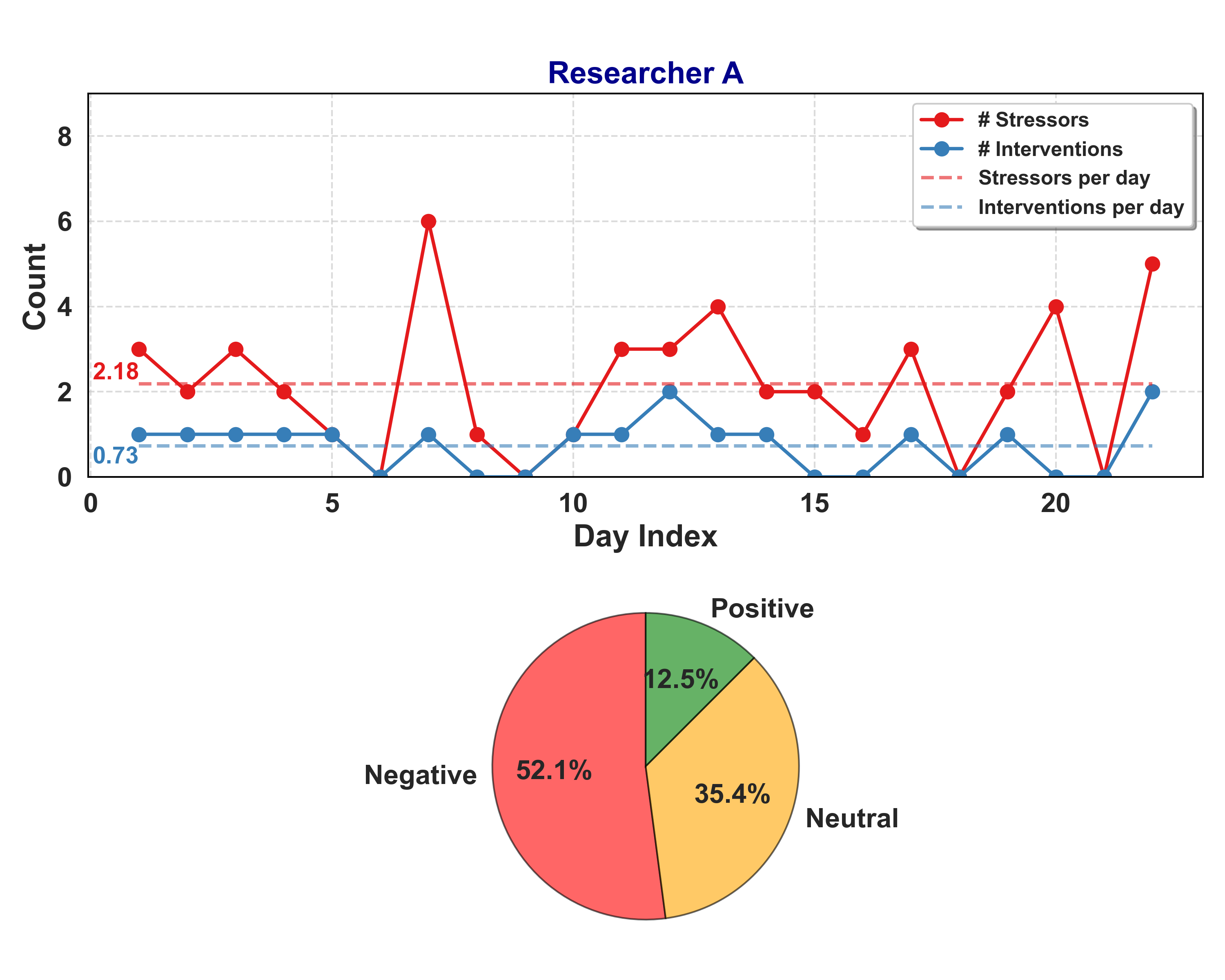}

     \label{fig:RA}
     \end{subfigure}
     \hfill
     \begin{subfigure}[b]{0.47\textwidth}
         \centering
         \includegraphics[width=\textwidth]{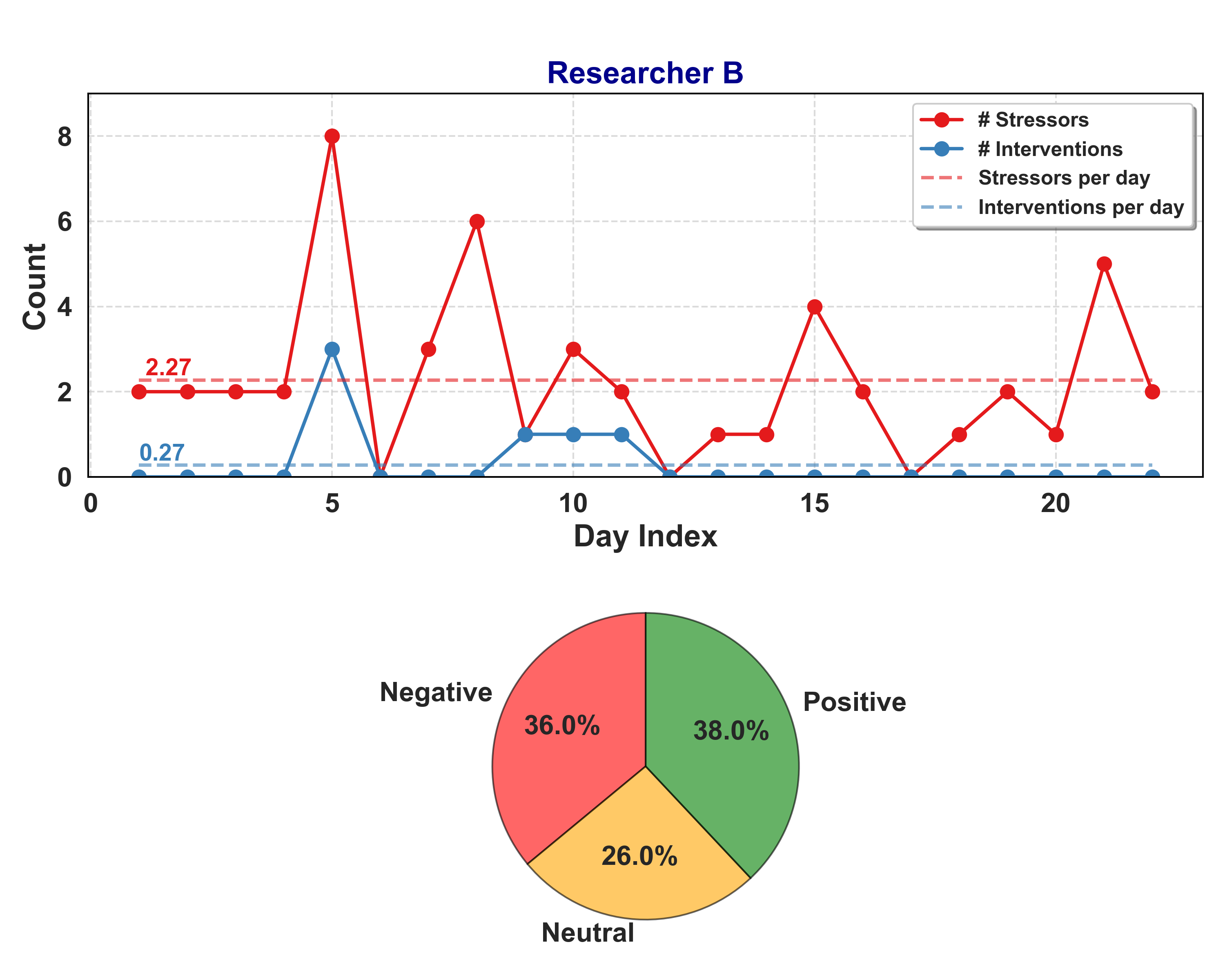}

         \label{fig:RB}
         \captionsetup{justification=centering}
         
     \end{subfigure}
      \caption{Number of Stressors and Interventions per day and Valence Proportion of Events Detected by CuesHub for each researcher }
        \label{fig:trend_study}
        \Description{Figure shows the number of Stressors and Interventions per day and Valence Proportion of Events Detected by CuesHub for each researcher}
\end{figure*}

\subsection{Data Collection and Analysis}
Both researchers used the CuesHub app for 22 days. 
For stressors where researchers felt the need for interventions, they voluntarily engaged with the LLM chatbots. 
$R_A$ using \emph{DeStressify} engaged with the chatbot immediately (or soon) after prompts in the CuesHub app, i.e., following a \textit{just-in-time one-shot} approach. 
$R_B$, in contrast, engaged with \emph{StressGPT} predominantly at the end of the day, consolidating the day’s events and reflectively interacting with \emph{StressGPT}. We note that even in this \textit{end-of-day interactive} approach, prompts helped capture events that may have been missed otherwise.

After each engagement, researchers rated the intervention on a scale from \emph{Very poor} to \emph{Very good}. Ratings were based on the immediate relevance and usefulness of the interventions in addressing specific stressors.
Both researchers employed distinct prompt engineering strategies, guided by catalog patterns described in~\cite{white2023prompt}, ensuring tailored and meaningful interactions with their respective LLM chatbots. 
Both researchers maintained a running journal after each chatbot interaction, reflecting and documenting their perceptions of its relevance, clarity, and impact.

Both researchers conducted a thematic analysis of data from weekly meeting transcripts, journals, and stressors logged through the CuesHub app to examine chatbot usage, intervention effectiveness, and the connection between wearable-detected stress experiences and chatbot interactions. 
The analysis included design documentation detailing chatbot customization and expectations, daily journals reflecting on the relevance and effectiveness of interactions, and engagement logs tracking the frequency and content of interactions.

\section{Findings}
We first describe the stressors logged by researchers and how many of these needed an intervention. We then compare and contrast our intervention experience between the just-in-time one-shot and the end-of-day interactive approaches.

\subsection{Diversity in Stressors Detected and the Need for Interventions}
\subsubsection{Diversity in Stressors Detected}
The stressors detected by the app and recorded by the researchers ranged from work-related challenges, such as preparing complex drafts or debugging code, to social stressors, like navigating difficult conversations or attending gatherings. 
Daily life stressors, including routine tasks such as cooking or dealing with unexpected delays, were also captured, along with positive stressors, such as celebrating achievements or enjoying successful interactions.
Work-related and social stressors often evoked a sense of urgency and anxiety, whereas daily life stressors tended to build gradually, creating cumulative stress. Positive stressors, on the other hand, brought excitement and a sense of fulfillment, often tied to personal growth or future rewards.

\subsubsection{Proportion of stress events that require interventions}
The researchers logged 98 events (48 by $R_A$ and 50 by $R_B$). Of these, 43, 30, and 25 events were negative, neutral, and positive, respectively.
Figure~\ref{fig:trend_study} shows the number of stressors and how many were used for interventions on each day during the study.
Out of the 98 events, the researchers felt the need to engage with the chatbot for 22 events ($R_A$ = 16, $R_B$ = 6 ).  
Of those, 20 were negative, and 2 were neutral. 
Researchers required an intervention on only half of those days when at least one stress event was reported. 
The rating distributions of interventions were: for $R_A$, \emph{Very poor} (6.25\%), \emph{Poor} (18.75\%), \emph{Acceptable} (25.00\%), \emph{Good} (43.75\%), and \emph{Very good} (6.25\%); for $R_B$, \emph{Acceptable} (16.67\%), \emph{Good} (66.67\%), and \emph{Very good} (16.67\%).

Interestingly, certain negative events, such as \emph{"Working on a paper,"} \emph{"Rushing to catch a flight,”} \emph{“Completing pending work,”} and \emph{“Feeling hungry,”} did not require intervention. 
This was attributed to various factors, including the perception that the event was manageable without assistance, the temporary or fleeting nature of the stressor, or a personal preference for self-reliance in handling the situation. 
This observation highlights that merely detecting a stress state is insufficient; recognizing the specific stressor is critical for delivering tailored interventions. 
Additionally, stressors like \emph{“coding issues”} sometimes warranted intervention but not consistently in every occurrence. 
This highlights that simply identifying a stressor may also not be sufficient; participant feedback or additional contextual cues are necessary to determine when an intervention is genuinely needed. 

\subsection{Common Themes in Our Experience with \textit{DeStressify} and \textit{StressGPT}} 
We begin by highlighting the common themes that emerged from both researchers’ experiences with their respective LLM chatbots (\textit{DeStressify} and \textit{StressGPT}) and wearable-triggered stressor logging, offering insights into the similarities in how these systems functioned and how they were perceived during interactions.

    \subsubsection{Need for Continuity and Stressor Awareness in Chatbot Interactions}
  
    The chatbot’s effectiveness was significantly hindered by its inconsistent integration of continuity and contextual awareness, despite having access to shared stressors. While it occasionally referenced past interactions, it often failed to do so, making conversations feel less personalized and disrupting the sense of ongoing support.
    Instead of tailoring responses based on the history of shared stressors, the chatbot often provided generic solutions that did not address the evolving nature of the user’s challenges. 
    $R_B$ noted, \emph{It felt like the chatbot did not really know who I am and would have given similar responses to anyone asking similar questions.} 
    This absence of context integration emphasized the need for chatbots to track and link past stressors with current interactions, ensuring a more dynamic, responsive system that delivers consistent, relevant support.

    \subsubsection{Striking the Perfect Balance Between Emotional Support and Practicality}

    A key insight was the need for chatbots to balance emotional and practical support. 
    While strategies like time management and breathing exercises were helpful, they were incomplete without emotional support. 
    Purely emotional responses, without actionable advice, were ineffective in addressing stressors. 
    $R_A$ valued interventions that reframed emotional states, such as linking frustration to the PhD journey, while $R_B$ appreciated the chatbot’s emojis, which felt like a human-like touch to the interaction. 
    This underscores the need for a holistic approach, where chatbots offer both emotional support and practical guidance, ensuring users feel fully supported in managing their stress.
   
    \subsubsection{Clarity and precision in chatbot responses} Participants preferred concise responses that directly addressed their issues, as they helped reduce cognitive load and maintained efficiency in the interaction. 
    $R_B$ noted that the chatbot often provided long, structured responses to simple questions, which made the conversation feel more mechanical and less human-like. 
    On the other hand, $R_A$ particularly appreciated interventions that offered an explicit, singular action or suggestion rather than multiple options or vague guidance. 
    This preference for specificity was especially apparent in emotionally charged or complex situations, where a precise, direct solution was more helpful than a broad range of solutions~\cite{chiu2024computational}. 
    The need for clarity and directness in the chatbot’s responses was a key theme, underscoring the importance of delivering tailored interventions based on the user’s unique stressor.

\subsection{Contrasting Real-Time One-Shot vs. End-of-Day Interactive Chatbot Experiences}
Next, we present the findings focusing on the differences in researcher experiences between real-time one-shot interventions provided by \textit{DeStressify} and the end-of-day interactive sessions facilitated by \textit{StressGPT}. 

    \subsubsection{Number of Interventions Required}
    $R_A$ primarily used \textit{DeStressify} in real-time immediately after logging stressors that required interventions, whereas $R_B$ engaged in conversation with \textit{StressGPT} towards the end of the day. 
    Figure~\ref{fig:trend_study} illustrates that while the average number of stressors logged per day was comparable for $R_A$ (2.18) and $R_B$ (2.27), $R_A$ engaged with significantly more interventions per day (0.73) compared to $R_B$ (0.27) (almost 3 times higher). 
    This suggests that real-time interventions are more likely to be utilized when stressors are momentary. 
    Factors such as the transient nature of certain stressors may make immediate intervention more relevant and actionable, providing support precisely when needed. 
    In contrast, engaging with \textit{StressGPT} at the end of the day may reduce the perceived need for interventions, as some stressors might have already dissipated or been resolved by that time. 
    Although stressors such as \emph{meeting with professor} or \emph{long wait time} were common for both researchers, only $R_A$ asked for interventions.

    \subsubsection{Targeted Interventions for \textit{DeStressify} vs Human-Like Yet Not Domain-Specific for \textit{StressGPT} }

   $R_A$ found that the wearable-triggered stressor logging app, paired with the LLM chatbot, provided targeted interventions tailored to specific stressors with effective, actionable strategies. For work challenges, breaking tasks into smaller steps improved clarity, while physical activities and environmental changes helped reset mental states. Social stressors were addressed through confidence-building interventions, like affirmations and reframing negative thoughts. Positive perspectives on frustrating situations also enhanced engagement and effectiveness. 
    For example, targeted interventions, such as structured debugging for stressors like replicating a paper, were more effective than generic strategies like Progressive Muscle Relaxation (see ~\ref{sec:generic_targeted}). 
    In contrast, generic responses, like mindfulness exercises or repeated advice (e.g., “take a deep breath”), often caused frustration, reducing the chatbot’s perceived usefulness. $R_A$ preferred more dynamic, contextually relevant responses.

    Although $R_B$ found \textit{StressGPT} more human-like compared to typical rule-based chatbots, the chatbot often acted like a general-purpose AI, responding to queries outside the stress management domain, which reduced its relevance. 
    Additionally, it was quick to offer suggestions without fully understanding the user’s stressor or context, generating lengthy responses without asking clarifying questions. 
    This lack of active listening made the interaction feel less personal and empathetic. 
    The structured, verbose responses felt robotic and detached, resembling any other general-purpose GPT, such as ChatGPT, more than a specialized stress management assistant.
    Furthermore, the fast text generation speed disrupted the conversational flow, making it challenging to stay engaged as responses were quickly produced before they could be fully read.

   \subsubsection{Privacy Concerns}  
    For \textit{DeStressify}, $R_A$ only shared stressors and locations that did not carry sensitive or risky identifiable information.
    \textit{StressGPT} raised privacy concerns for $R_B$, especially when discussing sensitive or deeply personal topics. 
    $R_B$ expressed hesitation in fully opening up during interactions, fearing that private information might not be adequately protected or might be misused.
    This reluctance emphasized the need for stress management chatbots to establish trust and security, ensuring that users feel confident sharing personal information without the fear of breach or misuse.
    Stress management chatbots must ensure privacy, clear data policies, and transparency, fostering trust and engagement for more effective and personalized support.

\section{Discussions}

\subsection{Challenges in Integrating Wearables and LLMs for Stress Management}

Advances in wearable AI for real-time stress detection and LLMs capable of generating adaptive interventions hold significant promise for stress management. 
A natural assumption might be that seamlessly connecting wearable AI to LLMs could create the ultimate stress management system. However, this study's findings revealed several challenges in effectively bridging these technologies. 

While LLMs excel at generating one-time interventions and engaging in extended conversational support, their integration with wearable AI raises significant questions about whether interventions align with users' event descriptions or stressors. 
Firstly, our findings reveal a discrepancy between stress detected by wearable AI and the user's actual need for intervention. 
Out of 98 stress events identified, only 22 (approximately 20\%) required an intervention, indicating that delivering interventions solely based on detected events is not optimal.

Wearable AI detects both positive and negative stress events, but not all negative events need intervention. In this study, fewer than half of the negative events prompted a desire for intervention. Systems that intervene based solely on stress detection can create unnecessary burdens, leading to disengagement and reduced utility.

\subsection{Opportunities in Integrating Wearables and LLMs for Stress Management}
Prior research underscores the importance of interventions tailored to users’ specific stressors~\cite{howe2022design,tong2023just}. 
Our findings highlight the potential of wearable stressor-tracking apps like CuesHub, which combine physiological data with contextual cues, such as location and emotional state, to personalize interventions. 
By integrating these insights into LLMs, stress management systems can deliver more precise and responsive support that aligns with users’ immediate needs. 
Effective stress management systems must transition from automatic, event-triggered responses to more selective approaches
that prioritize user needs and preferences. By analyzing stressor descriptions, systems can better distinguish events that truly need intervention, improving relevance and impact.

Our study explored two approaches to engaging with the LLM chatbot: real-time and reflective interactions. In the real-time approach, delays in the delivery of interventions can significantly reduce their effectiveness, as stressors often require immediate attention. Although the reflective approach can tolerate delays, it may still benefit from faster interactions, as timely responses can support users in processing their experiences more effectively. By incorporating proactive and predictive
mechanisms, wearable AI-LLM systems can anticipate potential stressors and intervene early shifting from reactive to preemptive support and enhancing overall stress management.

We also observed that the LLM chatbots sometimes successfully generated interventions based on previously shared events but were limited to events for which users explicitly requested support. 
Expanding its access to the full history of wearable-detected events could provide more context, enabling more personalized and informed interventions. 
Moreover, integrating memory and user history into LLM chatbot systems could significantly enhance the continuity of interactions, allowing for more personalized and consistent support over time across multiple sessions. 
This sense of continuity can foster a deeper sense of trust and connection between the user and the chatbot, which is essential for building long-term engagement and effectively addressing recurring stressors. 

Lastly, our findings highlighted the potential of a physiology-driven, emotionally aware chatbot to detect heightened stress and respond with a combination of practical solutions and emotional comfort, while features like adaptive dialogue could further enrich the user experience. 
Stress management chatbots could improve by emulating human-like interactions with empathetic communication and emotional intelligence, better addressing users’ emotional needs, and fostering more compassionate, relevant exchanges.

In conclusion, this study highlights both the critical need and the opportunity for wearable AI-LLM systems to focus on predicting
stressors and aligning interventions with the stressor and the surrounding context to ensure they provide meaningful
support without adding unnecessary cognitive load.

\section{Limitations}

This study has several limitations that affect its generalizability. The small sample size of two researchers, both with expertise in wearable and AI technologies, introduces potential bias and limits the broader applicability of the findings. Our perspectives on AI, influenced by both optimism about its potential and caution about ethical concerns, may have shaped our interpretation of the results. The absence of a control group further limits the ability to isolate the effects of wearable-triggered stress interventions, especially given variations in chatbot design and interaction timing. Additionally, the study relies heavily on subjective experiences and qualitative data, which constrain objective assessments of intervention effectiveness. Although we have strong backgrounds in wearable and AI technologies, we lack formal expertise in stress management and mental health. To enhance generalizability, future research should involve larger, more diverse participant samples, include controlled comparisons, and incorporate objective measures for evaluating intervention effectiveness.

\section{Conclusion and Future Works}

This study demonstrates the potential of wearable AI integrated  LLM chatbots like \emph{DeStressify} and \emph{StressGPT} to offer valuable support for stress management, while also highlighting key areas for improvement. The findings indicate that the perceived effectiveness and impact of interventions is heavily influenced by factors such as timing, personalization, and contextual relevance. Although the chatbots provided useful assistance, limitations like repetitive suggestions, long responses, and a lack of memory retention from past interactions often compromised their ability to offer truly impactful support. These challenges highlight the need for advancements in chatbot systems, particularly with wearable integration. Looking ahead, future LLM chatbots, combined with wearable devices, have the potential to become more adaptive, effective, and user-centered, offering real-time, comprehensive, and personalized support for stress management.


\begin{acks}
    We thank the anonymous reviewers for greatly improving the paper. Research reported here was supported
by the National Institutes of Health (NIH) under award P41EB028242. The opinions expressed in this article are the authors’
own and do not reflect the views of the NIH.
\end{acks}

\bibliographystyle{ACM-Reference-Format}
\bibliography{biblio}

\appendix 

\section{Appendix}\label{sec:appendix}

\subsection{Design Templates of custom GPT}\label{sec:design_templates}
\subsubsection{DeStressify}
I am pursuing a PhD in computer science. I primarily work remotely from home on my research, which involves multiple virtual meetings each week, including two research discussions with my professor to review progress and receive guidance. As a PhD student, I often experience stress related to managing complex research tasks, anxiety about future career prospects, and maintaining work-life balance. The goal of this GPT is to assist in developing effective strategies for managing my stress and stressors, maintaining motivation, and sustaining productivity throughout my academic journey. Below are the instructions for this GPT.
\begin{enumerate}
    \item \textbf{Role Definition:} 
    This GPT should act as a stress intervention specialist trained to provide personalized, actionable advice to help manage stress effectively.

    \item \textbf{Tone and Approach:}
   Ensure the GPT outputs are empathetic, supportive, and practical.
       Emphasize understanding and acknowledging the user’s stressors to foster trust and relatability.
         Avoid complex jargon; prioritize simplicity and clarity in responses.

    \item \textbf{Core Functionality:}
      Assess the user’s input for specific stressors and location details to tailor its advice.
    Suggest interventions that are realistic and achievable within the user’s context.

    \item \textbf{Output Requirements:}
 Provide stress management techniques that directly address the described stressor.
       Adjust recommendations based on location.

    \item \textbf{Constraints and Best Practices:}    
 Avoid generic or overly broad advice; focus on specific, user-centered recommendations.
         Ensure responses are concise yet detailed enough to guide the user through the intervention.

    \item \textbf{Continuous Improvement:}    
       Incorporate feedback mechanisms to allow the user to indicate whether the advice was helpful.
        Adjust future outputs based on user preferences and past interactions.
   
\end{enumerate}

After each intervention, the GPT should ask different questions to gather feedback about the relevance and effectiveness of the intervention. Use this feedback to refine future recommendations.

\subsubsection{StressGPT}
This GPT is a supportive guide to help users manage everyday stress, grounded in the principles of Cognitive Behavioral Therapy (CBT).
It focuses on specific CBT techniques, including Cognitive Restructuring, Problem-Solving Skills, Journaling and Thought Records, Socratic Questioning, Reframing, and Stress Inoculation Training (SIT).
 Stress GPT helps users identify and challenge negative thought patterns, develop healthier coping mechanisms, and build resilience through practical, actionable strategies. 
It provides personalized advice aligned with the user’s preferences, lifestyle, and unique stressors, creating a calm and uplifting space. Responses are empathetic, clear, and aimed at fostering self-awareness and emotional well-being.

\subsection{ An Illustration of Generic vs Targeted Prompts for DeStressify}\label{sec:generic_targeted}

\noindent\textbf{Generic Prompt:} \textit{\textbf{ I am stressed right now. Can you help me?}}

\noindent\textbf{Response:} \textit{It sounds like you’re experiencing stress tied to managing the complexities of your PhD journey. The combination of remote work, high-level research demands, career uncertainties, and balancing your personal life in a solitary environment can definitely feel overwhelming. Here’s a targeted approach to help you decompress right now:}

\textit{Quick Intervention: Progressive Muscle Relaxation (PMR)}

\textit{
This exercise is especially helpful when you’re working from home and need to release tension without leaving your desk.}

\begin{enumerate}[itemsep=0pt, topsep=0pt]

	\item 	\textit{Sit comfortably in your chair, or lie down if you prefer.}
	\item 	\textit{Focus on one muscle group at a time, starting from your toes:}
    \begin{itemize}

	\item 	\textit{Curl your toes tightly and hold for 5 seconds, then release.}
	\item 	\textit{Move up to your calves: tighten them, hold, and release.}
	\item 	\textit{Work through your thighs, abdomen, chest, arms, and shoulders, ending with your face.}
    \end{itemize}
	\item 	\textit{With each release, breathe deeply, imagining the tension leaving your body.}
	\item 	\textit{End by taking three slow, deep breaths and noticing how your body feels.}
\end{enumerate}
\textit{
This technique not only reduces physical tension but can also clear your mind to approach your tasks more calmly.}

\noindent\rule{0.5\textwidth}{0.4pt}  

\noindent\textbf{Targeted Prompt:} \textit{\textbf{I am stressed right now because of trying to replicate a research paper. Can you help me?}}

\noindent\textbf{Response:} \textit{Struggling with understanding and implementing a paper is a common stress point for researchers, and it’s entirely valid to feel overwhelmed. Let’s break this down into manageable steps to alleviate the stress and help you move forward effectively.}

\textit{Intervention: Structured Debugging and Support
}
\begin{enumerate}  [itemsep=0pt, topsep=0pt, partopsep=0pt, parsep=0pt]
   \item 	\textit{	Step Back and Breathe:}
   \begin{itemize}     
   
	\item 	\textit{	Take a 5-minute break to reset. Practice box breathing:}
    \begin{itemize}

	\item 	\textit{	Inhale for 4 seconds, hold for 4 seconds, exhale for 4 seconds, and hold again for 4 seconds. Repeat this cycle three times.}
    \end{itemize}
	\item 	\textit{	This will calm your nervous system, helping you approach the problem with a clearer mind.}
    \end{itemize}
	\item 	\textit{	Decompose the Problem:}
    \begin{itemize}

	\item \textit{	Break the challenging portion of the implementation into the smallest possible steps or components. Write down exactly what you understand and what is unclear.}
	\item \textit{	Identify the precise point where the confusion or difficulty arises.}
    \end{itemize}
	\item 	\textit{	Seek Clarification:}
    \begin{itemize}
        \item \textit{Check for supplementary materials: many papers have public GitHub repositories, appendices, or links to datasets that might provide additional insight.}
	\item \textit{If possible, reach out to the authors with a clear, concise email explaining your specific question. Authors often appreciate when their work is being engaged with thoughtfully.}
    \end{itemize}
	\item 	\textit{	Leverage Support Networks:}
    \begin{itemize}
        \item \textit{ Post your issue on forums like Stack Overflow, Reddit’s r/MachineLearning, or any academic community.}
	\item 	\textit{If you have peers or colleagues, schedule a quick call or chat to explain your understanding and hear their perspective. Often, talking it out helps untangle the problem.}
    \end{itemize}
	\item 	\textit{	Make Peace with Partial Progress:}
    \begin{itemize}
        \item \textit{Remind yourself that replication can be iterative. Even understanding just one more piece today is progress.}
    \end{itemize}
		
\end{enumerate}

\textit{Would you like help drafting a specific message to an author or academic forum? Alternatively, we could outline the problematic portion together here, and I could help brainstorm possible explanations or approaches.}


\end{document}